\documentclass[aps,twocolumn,showlabels,showrefs,amsmath,amssymb,pre,superscriptaddress, floatfix, colors]{revtex4}

\usepackage{graphicx}
\usepackage{dcolumn}
\usepackage{bm}
\usepackage{amssymb}
\usepackage{hyperref}
\usepackage{multirow}
\usepackage{xcolor}

\begin{document}
\title{Linear response theory and Green-Kubo relations for active matter}
\author{Sara Dal Cengio}
\affiliation{Departament de F\'isica de la Mat\`eria Condensada, Universitat de Barcelona, Mart\'i i Franqu\`es 1, E08028 Barcelona, Spain}
\author{Demian Levis}
\email[  ]{demian.levis@epfl.ch}
\affiliation{Departament de F\'isica de la Mat\`eria Condensada, Universitat de Barcelona, Mart\'i i Franqu\`es 1, E08028 Barcelona, Spain}
\affiliation{CECAM Centre Europ\'een de Calcul Atomique et Mol\'eculaire, \'Ecole Polytechnique F\'ed\'erale de Lausanne, Batochime, Avenue Forel 2, 1015 Lausanne, Switzerland}
\affiliation{UBICS University of Barcelona Institute of Complex Systems, Mart\'i i Franqu\`es 1, E08028 Barcelona, Spain}
\author{Ignacio Pagonabarraga}
\affiliation{Departament de F\'isica de la Mat\`eria Condensada, Universitat de Barcelona, Mart\'i i Franqu\`es 1, E08028 Barcelona, Spain}
\affiliation{CECAM Centre Europ\'een de Calcul Atomique et Mol\'eculaire, \'Ecole Polytechnique F\'ed\'erale de Lausanne, Batochime, Avenue Forel 2, 1015 Lausanne, Switzerland}
\affiliation{UBICS University of Barcelona Institute of Complex Systems, Mart\'i i Franqu\`es 1, E08028 Barcelona, Spain}

\begin{abstract}
We address the question of how interacting active systems in a non-equilibrium steady-state respond to an external perturbation. We establish an extended fluctuation-dissipation theorem for  Active Brownian Particles (ABP) which highlights the role played by the local violation of detailed balance due to activity.  By making use of a Markovian approximation we derive closed Green-Kubo expressions for the diffusivity  and mobility of ABP and quantify the deviations from the Stokes-Einstein relation. We compute the linear response function to an external force using unperturbed simulations of ABP and compare the results with the analytical predictions of the transport coefficients.   Our results show  the importance of the interplay between activity and interactions in the departure from  equilibrium linear response. 
\end{abstract}

\maketitle

Linear response theory describes how a (small) external perturbation affects the macroscopic properties of a system \cite{MarconiReview}. If the unperturbed system is initially in equilibrium, its response to small perturbations is generically related to  equilibrium fluctuations through the fluctuation-dissipation theorem (FDT).
In this situation,  one can derive  exact expressions for the transport coefficients in terms of equilibrium correlations, the so-called Green-Kubo relations, which are among the very rare general results in non-equilibrium statistical mechanics \cite{KuboBook}. 
However, intrinsically non-equilibrium systems, such as active matter, lie beyond the scope of this framework. Indeed, active matter stands for systems made of components which typically convert energy from their environment into motion in a way that breaks detailed balance \cite{ReviewMarchetti, ReviewCates, ReviewFodor, battle2016broken}.
Therefore, equilibrium states cannot be considered as the reference unperturbed states in that case. 
%

Extensions of the FDT to non-equilibrium states have been recently derived  
\cite{BaiesiMaesPRL, BaiesiMaesNJP, ChetriteFDT, ProstJoanny, SeifertSpeck, SpeckSeifert}. However, these approaches have not been applied to locally driven active systems and general Green-Kubo-like expressions, relating their transport coefficients with non-equilibrium steady-state correlations, have not been established yet. 
Establishing the nature of these relationships has deep consequences because of the insight they provide in the nature of non-equilibrium response in active matter. For example, experiments on microswimmer suspensions have revealed interesting rheological behaviour \cite{BacteriaClement, RafaiPRL2010} and rectification phenomena in the presence of asymmetric boundaries \cite{diLeonardoWheel,KoumakisDiLeonardo, SamuelWheel}, opening up the possibility to exploit the non-equilibrium character of active matter to control transport at the microscale and extract energy from it. Harvesting the potential of these systems thus needs the development of an extended response theory: 
This is the overall aim of the present work. 


We start by considering a generic
 time evolution of the probability $\Psi(\Gamma,t)$ to find the system in a state $\Gamma$:  
\begin{equation}\label{eq:1}
\frac{\partial \Psi}{\partial t} = \Omega \Psi = (\Omega_0 + \Omega_{\text{ext}}) \Psi\ .
\end{equation}
The  generator of the dynamics $\Omega$ can be split into an unperturbed, $\Omega_0$, and perturbed, $ \Omega_{\text{ext}}$, part. 
Using the operator identity $e^{\Omega t} = 1 + \int_{0}^{t} dt' e^{\Omega t} \Omega$, we can derive the following expression of the average of an observable $A$ at any time $t\geq 0$, $\langle A \rangle_t =\int d\Gamma \Psi(\Gamma,t) A(\Gamma)$, as
\begin{equation}\label{eq:2}
\langle A \rangle_t - \langle A \rangle_0 = \int_{0}^{t} ds \langle  \frac{\Omega_{\text{ext}} \Psi_0}{\Psi_0} A(s) \rangle_0 \, .
\end{equation}
Here $\langle* \rangle_0$ denotes averages over $ \Psi_0$, the steady-state initial distribution, verifying $\Omega_0\Psi_0=0$, while $A(s)=e^{\Omega^\dagger\,s} A(0)$ has been evolved with the full evolution operator (in the Heisenberg representation of the ensemble average). Note that, while the former expression is completely general and does not rely on any perturbation expansion (therefore also valid in the {non-linear  regime}), its application demands some knowledge of the distribution $\Psi_0$, or at least, the action of $\Omega_{\text{ext}}$ on it.  

For the  sake of clarity, let us focus first on an  equilibrium system of  $(N + 1)$   overdamped Brownian particles interacting by means of potential forces $\boldsymbol{F}_i=-\partial_i U$ ($\partial_i$ denotes the spatial gradient $\partial/\partial\boldsymbol{r}_i$) with diffusion coefficient $D_0$, mobility $\mu_0$ and inverse temperature $\beta=\mu_0/D_0$. In that case, Eq.~(\ref{eq:1}) corresponds to the Smoluchovski equation describing the time evolution of the  probability density of a point in configuration space  $\Gamma\equiv (\boldsymbol{r}_1, ...,\boldsymbol{r}_N,\boldsymbol{r}_n )$,  with $\Omega_0\equiv\Omega_{eq}=\sum_i \left[{\partial_i}\cdot ( D_0 {\partial_i} - \mu_0 \boldsymbol{F}_i )\right]$. 
 We perturb the system by applying a constant force $\boldsymbol{f}=f\boldsymbol{u}_x$  to a tracer particle described by $\boldsymbol{r}_n$, thus  $\Omega_{\text{ext}}=-\mu_0\boldsymbol{f}\cdot \partial_n$ and 
\begin{equation}\label{eq:B}
\Omega_{\text{ext}} \Psi_0=-\beta\mu_0\boldsymbol{f}\cdot\boldsymbol{F}_n \Psi_0\, .
\end{equation} 
From Eq.~(\ref{eq:2}) we find
\begin{equation}\label{eq:ittEq}
\langle A \rangle_t- \langle A \rangle_{0} = -\beta\mu_0 \boldsymbol{f} \cdot\int_0^{t}\mbox{d}s\langle \boldsymbol{F}_n e^{\Omega^{\dagger}s} A\rangle_0
\end{equation}
where $\Omega^{\dagger}$ is the adjoint (backwards) Smoluchovski operator \cite{Risken}. 
Eq.~(\ref{eq:ittEq}) constitutes a \emph{generalized nonlinear Green-Kubo expression}, relating a non-equilibrium average at time $t$ with an equilibrium time-correlation. 
Such approach, originally introduced in the context of glassy rheology, 
 is usually referred to as integration-through-transients (ITT) \cite{FuchsCatesPRL, fuchs2005, fuchs2009,ThomasITT, GazuzFuchs, SharmaBrader2016, ThomasABPMCT}. If we now choose $A\equiv F^x_n=\boldsymbol{F}_n\cdot\boldsymbol{u}_x$ in Eq.~(\ref{eq:ittEq}), and define the tracer mobility  $\mu\equiv \lim_{t\to\infty} {\langle \dot{\boldsymbol{r}}_n\rangle}/{\boldsymbol{f}}$, we find, in the linear regime $\boldsymbol{f}\to0$, the standard Green-Kubo relation: $\mu=\beta\int_{0}^{t} dt' \langle \dot{x}_n(t') \dot{x}_n(0) \rangle_0= \beta D$ (see \cite{SM} for a proof). 

We consider now Active Brownian Particles (ABP) \cite{LutzABP}, self-propelled with a constant velocity $v_0$ in the 2D plane along their orientation $\boldsymbol{n}_i=(\cos{\theta_i},\,\sin{\theta_i})$. The dynamic equations read
\begin{equation}\label{eq:ABP}
\dot{\boldsymbol{r}}_i(t)= \mu_0 \boldsymbol{F}_i+ v_0\boldsymbol{n}_i(t) + \boldsymbol{\xi}_i(t),\, \dot{\theta}_i(t)= \zeta_i(t)\, .
\end{equation}
where $\boldsymbol{F}_i$ accounts for all inter-particle potential forces (typically short-range repulsions) 
and $\mu_0$ is the mobility.
The  noise terms,  $\boldsymbol{\xi}_i$ and $\zeta_i$  are Gaussian and white, with zero mean and variance  $2D_0$ and $2D_{\theta}$, with $D_0=\mu_0/\beta$ the thermal Brownian diffusivity and $D_{\theta}$ the rotational diffusion coefficient, introducing the persistence time  $\tau=1/D_{\theta}$. Equilibrium is recovered both in the limit of  $v_0\to 0$ or $\tau\to 0$. 
 The ABP system  above is one of the reference models for active matter, and it has been studied extensively.  ABP has been characterized  in terms of thermodynamic quantities such as pressure and chemical potential \cite{Brady2014, solonPRL, WinklerPressure, JoanSoft, SolonThermo, paliwalChemical} and its phase behaviour analysed in great detail \cite{FilyMarchetti, RednerPRL, BialkeEPL,  FarageBrader, Siebert2017, LinoPRL}. Several recent works have studied its linear response  (and of similar active particles modelled in terms of an Ornstein-Uhlenbeck process) from different viewpoints: (i) introducing an effective temperature characterizing FDT violations \cite{LoiTeff, LevisTeff, szamelTeff, LeticiaFNL} ; (ii) taking equilibrium as the reference state and considering activity as a perturbation of it \cite{SharmaBrader2016, ThomasABPMCT, FodorPRL}; (iii) deriving  expressions of linear response functions 
in terms of weighted averages over the unperturbed dynamics \cite{szamelMalliavin, SolonResponse, LeticiaFNL},  in the same spirit as the Malliavin weights sampling \cite{WarrenAllenPRL}.  
In the present letter, we first characterize the violations of the FDT in ABP, showing how the non-equilibrium character of activity comes into play in an extended FDT that we establish, and then derive Green-Kubo expressions for its transport coefficients in terms of its non-equilibrium fluctuations. 

%

With the aim of characterizing transport in active matter,  we first analyze how ABP (in a non-equilibrium steady state) respond to an external perturbation.   
  Our starting point is the Smoluchowski operator corresponding to Eq.~(\ref{eq:ABP})
\begin{equation}\label{eq:SmolABP}
\Omega_0=\sum_i \left[{\partial_i}\cdot ( D_0 {\partial_i} - \mu_0 \boldsymbol{F}_i - v_0 \boldsymbol{n}_i ) + D_{\theta}  \frac{\partial^2}{\partial \theta_i^2} \right] 
\end{equation}
with a perturbation  
due to an  external force $\boldsymbol{f}=-\partial_n V$, where $V=-fx_n$, applied to a tagged particle located in $\boldsymbol{r}_n=(x_n,y_n)$.  A salient feature of ABP  is the absence of zero-flux steady-state solutions  of the Smoluchovski equation  (which can be seen from the impossibility to simultaneously satisfy the zero-current conditions for the angular and positional degrees of freedom). In the absence of odd variables under time-reversal, as it is the case for this overdamped dynamics, the absence of zero-flux solution is a necessary and sufficient condition for the violation of detailed balance \cite{GardinerBook}.  [In contrast to  Active Ornstein-Uhlenbeck Particles (AOUP) which fulfil detailed balance in the small drive limit  \cite{FodorPRL, Bonilla}]. 

Because the probability density must be positive we may write 
\begin{equation}\label{eq:ansatz}
\Psi_0 (\Gamma)\sim e^{-\beta\left[U (\{\boldsymbol{r}_i\})+ v_0 \tau \, \Xi (\{\boldsymbol{r}_i,\theta_i\})\right] }  \,
\end{equation} 
in terms of a generalized potential $v_0 \tau\,\Xi (\{\boldsymbol{r}_j,\theta_j\})$ 
 - related to the so-called  information potential 
 in stochastic thermodynamics  (see e.g. \cite{BaiesiMaesNJP}) -  encoding deviations from Boltzmann statistics.
The steady state condition $\Omega_0\Psi_0=0$ leads to 
\begin{equation}
 v_0 \partial_i \cdot (- \mu_0\tau\,\partial_i \Xi-\boldsymbol{n}_i)\Psi_0 + \tau^{-1} \partial^2_{\theta_i}\Psi_0=-\nabla_i\cdot \boldsymbol{j}_i
 \end{equation}
where $\nabla_i=(\partial_i, \partial_{\theta_i})$, introducing the steady-state local velocity
\begin{equation}
\boldsymbol{\nu}_i \equiv \boldsymbol{j}_i /\Psi_0=v_0 \left[ (\boldsymbol{n}_i + \mu_0\tau\,\partial_i \Xi ),  \beta\partial_{\theta_i}\Xi \right]\,.
\end{equation}
In order to make the connection between the steady-state current and the local violation of detailed balance explicit, we consider the time-reversed adjoint operator $\overline{\Omega}_0^\dagger$ (defined as $A(-t)=e^{\overline{\Omega}_0^\dagger t} A(0)$) and  find \cite{SM} 
\begin{equation}\label{eq:TRO}
\Omega_0^\dagger-\overline{{\Omega}}_0^\dagger=2\boldsymbol{\mathcal{V}}\cdot\boldsymbol{\nabla}
\end{equation}
where $\boldsymbol{\mathcal{V}}=\{\boldsymbol{\nu}_i\}$ and $\boldsymbol{\nabla}=\{\partial_i\,,\partial_{\theta_i}\}$ . 
Eq.~(\ref{eq:TRO})  shows  that violations of detailed balance and non-zero steady current are two faces of the same coin. In the passive limit $v_0 \to 0$ detailed balance is recovered and $\boldsymbol{\nu}_i\to0$.

Inserting Eq.~(\ref{eq:ansatz}) into Eq.~(\ref{eq:2}) yields
\begin{align}\label{eq:FDT}
&\langle A \rangle_t - \langle A \rangle_0 = \nonumber \\
&\beta\,  \left(\int_{0}^{t} ds\langle \, \dot{V}(0) A(s) \rangle_0 -  \int_{0}^{t} ds\langle\boldsymbol{\mathcal{V}}\cdot\boldsymbol{\nabla} V(0)  A(s) \rangle_0 \right) 
\end{align}
In the linear regime, the latter expression constitutes an extension of the FDT far-from-equilibrium \cite{BaiesiMaesPRL, BaiesiMaesNJP}. When $\boldsymbol{\mathcal{V}}=0$ one easily recovers the standard Kubo expression $\beta^{-1}R_A(t)= \frac{\partial}{\partial t} \langle x_n(0) A(t) \rangle_{eq} $
where $R_A(t)=\delta \langle A \rangle/\delta f$.
Activity is responsible for the second term $\propto \boldsymbol{\mathcal{V}}$ in Eq.~(\ref{eq:FDT}), which quantifies the local dissipation of energy required to maintain the nonequilibrium steady state, often referred to as housekeeping heat \cite{PhysRevLett.86.3463}.   

As opposed to the equilibrium FDT, the response of an active system is not completely determined by its fluctuations, but depends on the specific form of its steady-state distribution  \cite{Risken, SpeckSeifert, FodorPRL}. 
Our aim being to derive explicit Green-Kubo relations for ABP (which do not depend on an unknown generalized potential), we forbid ourselves to make any assumption about  steady-state properties but  \emph{only} rely on the \emph{dynamics}. Our starting point should thus be Eq.~(\ref{eq:ABP}), although for this  dynamics we cannot derive an analog of Eq.~(\ref{eq:B}). Thus, the ITT construction that leads us to Eq.~\ref{eq:ittEq} cannot be readily  followed. To overcome this difficulty we integrate out the angular variables and work with the following reduced dynamics \cite{FilyMarchetti, FarageBrader}
\begin{equation}\label{eq:RABP}
\dot{\boldsymbol{r}}_i(t)= \mu_0 \boldsymbol{F}_i+ \boldsymbol{\eta}_i(t) 
\end{equation}
where the noise $\boldsymbol{\eta}_i$ is approximately gaussian with zero mean and variance $\langle \boldsymbol{\eta}_i(t) \boldsymbol{\eta}_j(s) \rangle = (2 D_0  \delta(t-s)  + {v_0^2}e^{-  |t - s|/\tau}/2) \delta_{ij}\mathbf{1} $.
As usual, by integrating away some degrees of freedom one generates memory, here with a time-correlation $\tau\equiv 1/D_{\theta}$. 
 Even if in the non-interacting limit particles described  by Eq.~(\ref{eq:RABP}) diffuse at long times with a diffusivity $D_a = D_0 + v_0^2\tau/2$, 
 the difficulty resides on the non-Markovianity of the evolution, which cannot be formulated in terms of a Smoluchovski operator; a long standing problem  in statistical mechanics \cite{HanggiRev, Adelman, MaxiSancho, Fox}. At small correlation times, Fox developed a small$-\tau$ expansion that leads to an effective Smoluchowski equation and which has proven useful in the context of ABP and AOUP \cite{FarageBrader, Wittmann2017effective}. To lower order in this expansion, Eq.~(\ref{eq:RABP}) reduces to \cite{FarageBrader}
\begin{equation}\label{Foxeq}
  \Omega_0^{M} = \sum_{i=1}^{N} \partial_{i} \cdot \mathcal{D}_i(\Gamma) \left[ \partial_{i} - \beta \boldsymbol{F}^{eff}_{i} (\Gamma) \right] 
  \end{equation}
  where we have introduced an effective diffusivity  and interaction force
  \begin{align}\label{Fox}
 & \mathcal{D}_i(\Gamma) = D_0 + \frac{v_0^2\tau}{2} \left( 1 + \frac{{\tau\mu_0 \partial_i \cdot \boldsymbol{F}_i}}{1 - {\tau\mu_0 \partial_i \cdot \boldsymbol{F}_i}}\right)\\
 & \boldsymbol{F}^{eff}_i(\Gamma) = ( D_0{\boldsymbol{F}_i} - \beta^{-1}{\partial_i \mathcal{D}_i })/{\mathcal{D}_i}
  \end{align}
  [Note that  $\tau\mu_0 \partial_i \cdot \boldsymbol{F}_i<1$ must be ensured.] The dynamics encoded in Eq.~(\ref{Foxeq}) fulfils detailed balance. The non-equilibrium character of the problem is now encoded in the effective diffusivity (which now depends on the relative positions of all the particles) and forces (which do not derive from a potential). 
Although the steady-state distribution is non-Boltzmann, it  has now zero net current, thus lifting the difficulties associated with the operator Eq.~(\ref{eq:SmolABP}) and enabling us to proceed with the ITT construction. Indeed, if we consider the constant force perturbation 
we find 
\begin{equation}
\Omega^M\Psi_0=(\Omega^M_0+\Omega_{\text{ext}})\Psi_0=-\beta\mu_0\boldsymbol{f}\cdot\boldsymbol{F}^{eff}_n \Psi_0\, ,
\end{equation}
 which allows us to derive an analogue of Eq.~(\ref{eq:ittEq})
 \begin{equation}\label{eq:ittFox}
 \langle A \rangle_t-\langle A \rangle_{0} =-\mu_0 \beta\boldsymbol{f} \cdot\int_0^{t}\mbox{d}s\langle \boldsymbol{F}^{eff}_n e^{(\Omega^{M})^{\dagger}s} A\rangle_0
 \end{equation}
 
Eq. \ref{eq:ittFox} allows us to derive Green-Kubo expressions of the diffusivity and mobility which do not rely of the Stokes-Einstein relation, but only on the  time-evolution of the system under the Markovian approximation Eq.~(\ref{Fox}). By choosing $A\equiv F_n^x$ in Eq.~(\ref{eq:ittFox}) we get the following Green-Kubo relation for the mobility
 \begin{equation}
  \mu= \mu_0 \left( 1 - \mu_0 \beta \int_{0}^{\infty}ds \langle F^{eff, x}_n (0) F^{x}_n(s) \rangle_0 \right)
  \end{equation}
which, to first order in ${\tau \mu_0 \partial_i \cdot \boldsymbol{F}_i}$, reads
\begin{align}\label{eq:GKmu}
\frac{\mu}{\mu_0} = &1 - \frac{\mu_0^2}{D_a} \int_{0}^{\infty} ds \langle F_{n}^{x}(0) F_{n}^{x}(s) \rangle_0  \nonumber \\ 
&+ \frac{v_0^2  \tau^2 \mu_0^3}{2 D_a^2 } \int_{0}^{\infty} ds \langle F_n^{x}(0) \partial_n \cdot \boldsymbol{F}_n(0) F_n^{x}(s) \rangle_0   \nonumber \\ 
&+ \frac{v_0^2  \tau^2 \mu_0^2}{2   D_a} \int_{0}^{\infty} ds \langle \partial_n^{x} \partial_n \cdot \boldsymbol{F}_n(0) F_{n}^x (s) \rangle_0 
\end{align}
In equilibrium, only  the terms in the first line survive, capturing how interactions affect the ideal gas mobility. Here, activity plays a role in the statistics of collisions,  thus the force self-correlation function, and in the value of the pre-factor via the single particle diffusivity. The  remaining  two terms $\propto (v_0\tau)^2$ correspond to subdominant higher order correlations involving many-body interactions. 

To further characterize the departure from equilibrium linear response and, in that case, the Stokes-Einstein relation,  we make use of the expression (see \cite{SM} for a proof) 
 \begin{equation}
   \mu_0^2\langle {F}^x_n(0) \cdot {F}^x_n(t) \rangle_0=
 \langle \mathcal{D}_n \rangle_0\,\delta(t)-   \langle \dot{x}_n(0)  \dot{x}_n(t) \rangle_0
  \end{equation}
 which relates the force and the velocity autocorrelation functions, and whose functional form depends only on the properties of the evolution operator (the special case for equilibrium dynamics was derived in  \cite{Klein}). 
Once we identify the diffusivity with  the velocity self-correlation function  we  get the Green-Kubo expression 
\begin{equation}\label{eq:GKD}
D = D_a + \frac{\mu_0 v_0^2 \tau^2}{2  } \langle \partial_n \cdot \boldsymbol{F}_n \rangle_0  - \mu_0^2 \int_{0}^{\infty}ds \langle F^{x}_n(0) F^{x}_n(s)\rangle_0
\end{equation}
 which allows us to express Eq.~(\ref{eq:GKmu}) as
\begin{equation}\label{eq:StokesEinstein}
 \frac{\mu}{\mu_0} =  \frac{D}{D_a} - \frac{\mu_0 v_0^2\tau^2}{2   D_a} \langle \partial_n \cdot \boldsymbol{F}_n \rangle_0 + \text{h.o.t.}
\end{equation}
where h.o.t. refers to higher order terms (see Eq.~(\ref{eq:GKmu}))
 and which reduces to the usual Stokes-Einstein relation in the passive limit.  If inter-particle forces are divergence-free,   a modified Stokes-Einstein relation holds, $\mu =\beta_{eff}D$,  but with an effective temperature $k_BT_{eff}=D_a/\mu_0$ (see Eq.~(\ref{eq:StokesEinstein})).
For instance, in the dilute limit,  ABP behave as an equilibrium ideal gas at a higher $T_{eff}$   \cite{PalacciPRL, LevisTeff, szamelTeff, LeticiaFNL}].
 Indeed, genuine non-equilibrium behaviour (with no effective equilibrium description) results from the combined effect of interactions and activity,
  as observed in active colloidal suspensions \cite{Ginot} and proven for AOUP \cite{FodorPRL, StefanoPRX, Bonilla} (among other examples).  Incidentally, AOUP behave as an effective equilibrium system 
 when the potential has zero third derivatives, while in our case equilibrium-like response is recovered if the potential has zero second derivatives. 
 
 To illustrate our results and put them into test, we run particle based simulations of ABP Eq.~(\ref{eq:ABP}) with periodic boundary conditions. We consider the pair potential  $U(r)=(\sigma/r)^{12}$ and the following set of parameters: $\tau=0.02$, $\mu_0=1$ and $\beta=1$. We vary the Peclet number $\text{Pe}=v_0\tau/\sigma $ and the mean density $\phi=\frac{\pi \sigma^2N}{4 L^2}$ in a range for which the system remains in its homogeneous phase ($\text{Pe}\in(0\,:\,10)$ and $\phi\in(0.01\,:\, 0.2)$). 
 
  \begin{figure}[h]\label{fig}
\centering
\includegraphics[scale=0.47,angle=-90]{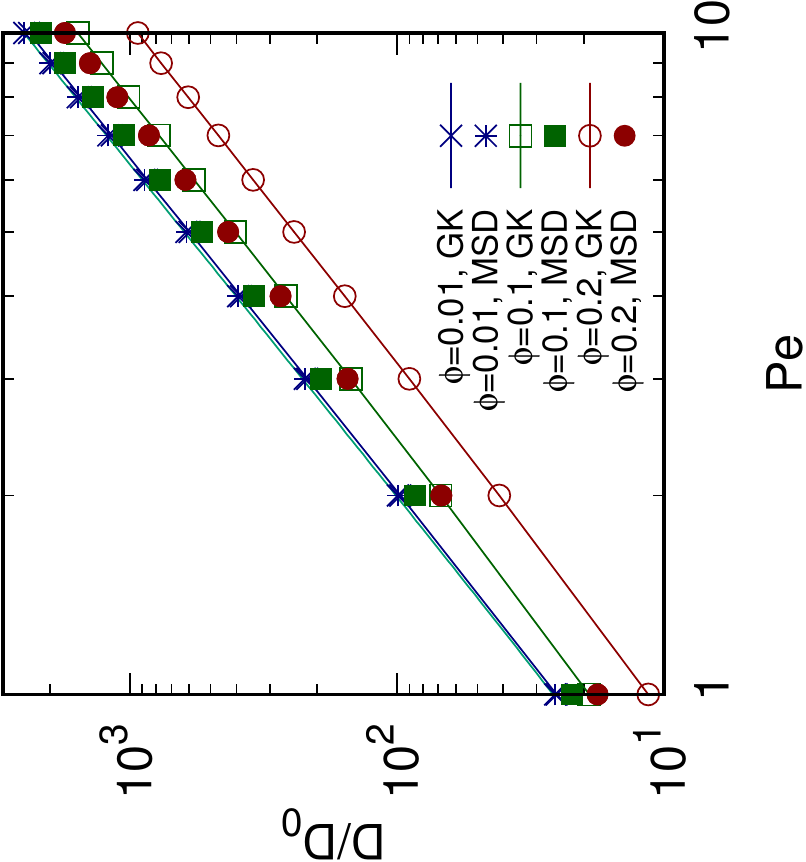}\includegraphics[scale=0.47,angle=-90]{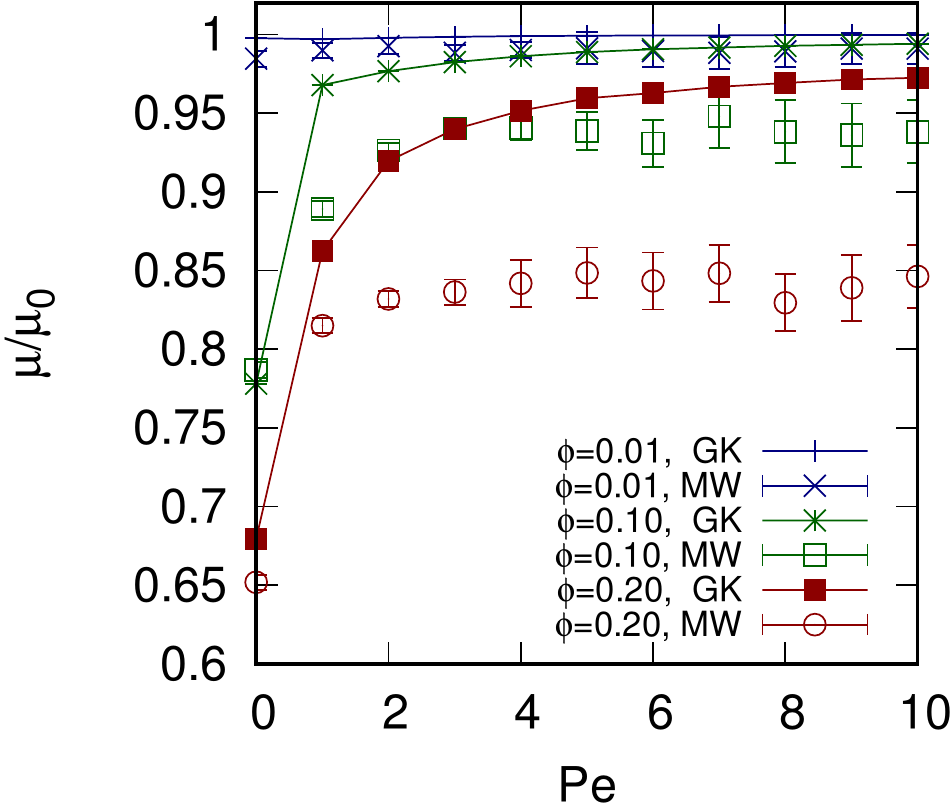}

\caption{Transport coefficients for  ABP  at $\phi=0.01,\,0.10$ and $0.20$ as a function of Pe.
Left: Diffusion coefficient obtained from the long time behaviour of the mean-squared displacement (MSD, with points) compared to the Green-Kubo expression eq. \ref{eq:GKD} (GK, with line-points). 
Right: Mobility obtained from simulations of  unperturbed ABP using Malliavin weights (MW, with error bars) compared to the Green-Kubo  prediction eq. \ref{eq:GKmu} (GK, with point-lines).  }\label{fig}
\end{figure}
 We  analyze the integrated response $\chi(t)=N^{-1}\lim_{f\to0}\sum_i^N(\langle x_i(t)- x_i(0)\rangle)/f$ of the particles' positions due to a constant force applied to all them  $\boldsymbol{f}_i=\epsilon_i f \boldsymbol{u}_x$, where $\epsilon_i=\pm 1$ with equal probability \cite{LevisTeff}.  We   compute $\chi$ using two different strategies: (i) we explicitly apply a small force  and measure the particle displacements it induces; (ii) we track the appropriate stochastic variables needed to compute the response function of interest  using simulations of the unperturbed dynamics.  The first 'direct method' involves computing displacements generated by a small perturbation that guarantees the linear regime.  The second 'Malliavin weight (MW) method'  overcomes the considerable numerical uncertainties (and cost) related to the control of a small perturbation parameter in Brownian dynamics simulations: we thus make an extensive use of it in the following. This method, originally introduced in the context of Monte Carlo simulations of spin systems \cite{Chatelain}, and then generalized to Brownian dynamics  \cite{WarrenAllenPRL,WarrenAllen}, was recently extended to active particle systems \cite{szamelMalliavin, LeticiaFNL}. 
 
We compute the diffusion coefficient $D=N^{-1}\lim_{t\to\infty}\sum_i\langle(\boldsymbol{r}_i(t)-\boldsymbol{r}_i(0))^2 \rangle/(4t)$ from the long-time behaviour of the mean-square displacement (MSD). 
 According to the MW method, the response function of interest can be expressed as \cite{WarrenAllenPRL,LeticiaFNL}
\begin{equation} \label{eq:MW}
\chi(t)=N^{-1}\sum_i \langle x_i (t) q_{i}(t)\rangle
\end{equation}
where $q_i$ is a Malliavin weight that evolves accordingly to 
$\dot{q}_{i}(t)=\sqrt{{\beta}/{2}}\sum_i  \xi^x_i(t)$,
and averages are taken over independent realizations of the unperturbed dynamics. 
We thus compute $\chi$ using Eq.~(\ref{eq:MW}) and extract $\mu=\lim_{t\to\infty}\chi(t)/t$ (after checking consistency with the 'direct method', see \cite{SM}). 
We also compute $D$ and $\mu$ using our Green-Kubo expressions. To be concrete, we compute from simulations of ABP, the different terms involving correlations and gradients of the potential that appear in Eq.~(\ref{eq:GKD}) and Eq.~(\ref{eq:GKmu}). 
The results obtained are shown in Fig.~(\ref{fig}).
The diffusion coefficient follows the same growth $\sim\text{Pe}^2$ as  the ideal gas in the parameter range explored, but decreases with $\phi$. Eq.~(\ref{eq:GKD}) underestimates the value of $D$ obtained   from the MSD but retains its functional dependence. 
Mobility is not affected by activity in the dilute regime and its value decreases as the density increases, as expected from Eq.~(\ref{eq:GKmu}) and previous works \cite{LevisTeff, LeticiaFNL}. At finite density and Pe, $\mu$ decreases with density but remains roughly constant for $\text{Pe}\gtrsim 2$. Such behaviour is reproduced by Eq.~(\ref{eq:GKmu}), although it overestimates the numerical value. 
Our Green-Kubo expressions predict the qualitative behavior of $D$ and $\mu$, despite we cannot reach a precise quantitative agreement at high Pe and $\phi$. Although such mismatch, expected from the basic assumptions behind the approximations made to reach Eq~(\ref{eq:GKD})  and Eq.~(\ref{eq:GKmu})(constrained to small values of $\tau$ and $\tau\mu_0 \partial_i \cdot \boldsymbol{F}_i$; neither too active nor too dense), prevents quantitative agreement, the derived Green-Kubo expression provides a general understanding on how the interplay between activity and interactions affects the transport properties of active particle systems.

The response of non-equilibrium systems is typically sensitive to  details of the unperturbed initial state, hence the difficulty to establish a general theory.
Such lack of universality is encoded in the presence of the generalized potential in extended fluctuation-dissipation relations, arising from the breakdown of detailed balance at the microscopic level. 
In this letter we set the stage for a systematic response theory of active systems obtained on pure dynamical grounds. Via a Markovian approximation scheme we overcome the aforementioned difficulty and we put forward a closed Green-Kubo expression for the mobility and diffusivity. This allows us to quantify the breakdown of the Stokes-Einstein relation due to the interplay between activity and inter-particle interactions. In presence of the latter, active systems have proven to substantially depart from the response of inanimate matter, contrary to the non-interacting limit for which an \textit{exact} equilibrium mapping exist.
Extending the present approach to other transport coefficients will then provide a theoretical framework to gain insights on the rheological properties of active matter \cite{BacteriaClement, RafaiPRL2010, diLeonardoWheel,KoumakisDiLeonardo, SamuelWheel}.

 \acknowledgments
 We warmly thank Thomas Voigtmann, Udo Seifert, Roland Netz, Jose M. Sancho and Miguel Rubi for discussions and suggestions. 
 SDC and IP acknowledge funding from the European Union's Horizon 2020 program under ETN grant agreement 674979-NANOTRANS.
DL acknowledges funding from EU Horizon 2020 program under the Marie Sklodowska-Curie Actions H2020-MSCA-IF grant agreement no. 657517. 

\bibliographystyle{unsrt}
\bibliography{biblio.bib}
\end{document}